# Disorder, defects and bandgaps in ultra thin (001) MgO tunnel barrier layers

P. G. Mather, J. C. Read, R. A. Buhrman

Cornell University, Ithaca, NY 14853-2501

## Abstract

We report scanning tunneling spectroscopy studies of the electronic structure of 1.5 to 3 nm (001) textured MgO layers grown on (001) Fe. Thick MgO layers exhibit a bulk-like band gap, ~ 5-7 eV, and sparse, localized defect states with characteristics attributable to oxygen and, in some cases, Mg vacancies. Thin MgO layers exhibit electronic structure indicative of interacting defect states forming band tails which in the thinnest case extend to ~ ±0.5 V of the Fermi level. These vacancy defects are ascribed to compressive strain from the MgO/Fe lattice mismatch, accommodated as the MgO grows.

# I Introduction

The properties of thin insulator layers are crucial to the performance of many electronic systems and devices, and in recent years major theoretical advances have been made in the calculation of the electronic behavior of systems that include such layers. However, in general such calculations must assume an ideal insulator structure while real insulating systems almost invariably contain defects that, even if only present in a limited density, can dominate the overall behavior of the system. One such system that is currently subject to intense research activity is the magnetic tunnel junction (MTJ), which is on the verge of becoming the enabling component in advanced magnetic hard drives and non-volatile magnetic memories. Essential to realizing this breakthrough is the achievement of robust, reproducible MTJs with the highest possible tunneling magnetoresistance (TMR) at the optimum specific-resistance level, which varies with application. A seminal scientific development in this field has been the theoretical prediction [1,2] of extremely high TMR in MTJs utilizing (001) oriented MgO barrier layers epitaxially formed on (001) Fe electrodes, followed by experimental realizations of high TMR with MgO MTJs utilizing both epitaxial and textured Fe and Fe alloy electrodes [3-5]. Results to date, however, are considerably below theoretical predictions, especially in the ultra-thin barrier limit, and fitting of current-voltage (*I-V*) characteristics indicate tunnel barrier heights ~0.5 eV, far lower than indicated by ideal models, and well below what would be expected from the results of scanning tunneling spectroscopy (STS) studies of carefully grown epitaxial MgO layers on single crystal Ag and Fe surfaces [6,7]. These results have been attributed to vacancies or other structural defects within the barrier [4,5] that arise from imperfect lattice matches and nonideal growth conditions. Here we report STS studies of the electronic structure of such imperfect MgO layers, which reveal how, as disorder in the barrier material varies, either as the result of thermal processing or thickness variation, the MgO bandgap, $E_G$, can change markedly, from a bulk-like value, > 7 eV to ~ 1 eV in the thinnest, most disordered case. Growing barrier layers via a method that results in both O and Mg vacancies, and hence in the layers being somewhat responsive to moderate thermal annealing, allows us to examine isolated vacancy-defects and find that the former results in electronic states at ~ 1 eV and 2.25 eV below the MgO conduction band minimum (CBM), while the latter yields a localized state ~ 1 eV above the valence band maximum (VBM). At high defect densities these localized states and the resultant on-site disorder in the oxide produce defect bands and band-tails that extend to within 0.5 eV of the Fermi level of the underlying Fe electrode.

## II Experimental Proceedures

We prepared the samples used in this study in an interconnected ultra-high-vacuum chamber containing an annealing oven, deposition sources, and a scanning tunneling microscope (STM) for *in-situ* analysis. We used GaAs substrates to provide a good lattice match for the growth of Fe (1.2% mismatch). First, we etched the surface oxide in an $NH_4OH$ solution and then vacuum annealed (610° C) the substrates to further clean and remove the thin oxide layer resulting from brief atmospheric exposure. Samples were then cooled to ~150° C to prevent FeAs formation. We deposited 4 nm of Fe via electron beam (e-beam) evaporation at a rate of 3 nm/ min and a pressure of $1 \times 10^{-9}$ Torr. X Ray Diffraction (XRD) measurements show this produces high quality base electrodes with a FWHM of 0.8° in 2Θ, and 0.3° in χ, indicating strongly (001) oriented films. Next, we deposited MgO by either e-beam or a two-step DC magnetron sputtering process. E-beam samples were deposited from a single crystal MgO source at a rate of 3 nm / min, and a pressure of $1 \times 10^{-8}$ Torr. A residual gas analyzer showed that this increased pressure was excess oxygen, indicating the deposited film is oxygen deficient. Sputtered samples were produced by first sputtering from a Mg target in 3mTorr Ar, and then switching to a reactive oxygen sputtering mode to deposit MgO and oxidize the initial Mg layer. XRD measurements show a FWHM of 0.8° (1.7°) in 2Θ and 2° (4°) in χ, for evaporated (sputtered) films, indicating good texturing of the barrier layer. The samples were then transferred into the STM for *in-situ* measurement at room temperature, some after being vacuum-annealed at ~375 C for 1 hour. We employed a lock-in amplifier to acquire spatially resolved DOS simultaneously with the topographic data and to provide direct $dI/dV$ measurements during spectroscopic acquisition.

### III Data and Interpretation

Figure 1 shows topographic (a) and DOS ($dI_t/dV_t$) (b) images from a thin, sputtered MgO layer (5 Å Mg / 10 Å MgO) imaged with a tip bias $V_t = 2$ V and tunnel current $I_t = 100$ pA after annealing at 375 C. The 2 V DOS map indicates that there are substantial variations in the electronic structure across the surface of the MgO layer, but with no clear correlation with the MgO topography. DOS measurements ($dI_t/dV_t$ vs. $V_t$), as shown for example in Fig. 1 (c), reveal that this MgO layer exhibits a band-gap that is rather symmetric about the Fermi level ($V_t = 0$) and that only varies marginally in amplitude from spot-to-spot from its average value of $E_G \sim$ 1.5 eV. Such measurements indicate that the variations in the DOS map (Fig.1a) result from local differences in the bias dependence of the band tails once the band-gap is exceeded. This is consistent with variations arising from where the Fermi level of the disordered MgO is pinned due to local fluctuations in defect character and density. Very similar DOS results (solid curve, Fig. 1c) are obtained from 5 Å Mg / 10 Å MgO layers examined prior to annealing, but in this case the symmetrical band-gap is only ~ 1.0 eV, *i.e.* in the case of the sputtered MgO layer, thermal annealing is found to increase $E_G$ by ~ 0.5 eV.

We find a substantial change in the DOS of the sputtered MgO layers when the thickness is increased by as little as 5 Å. In Fig. 2a, we show representative DOS measurements taken before and after annealing, which again reveal relatively symmetric band tails, but $E_G$ has increased with MgO thickness to ~ 1.5 eV before annealing and to ~ 2.5 eV afterwards. For still thicker MgO layers, 15 Å Mg / 15 Å MgO, the increase in $E_G$ continues, with Fig 2b showing representative DOS plots that yield $E_G$ ~ 6 eV before annealing, and nominally the same after annealing, although in some cases it is as high as ~ 7.5 eV (Fig. 3c), which is the value reported from STS measurements on well-ordered and annealed MgO layers epitaxially grown on Ag single crystal [7].

The behavior of the e-beam evaporated MgO layers is similar, but somewhat different in detail. For a 15 Å e-beam layer, DOS measurements again show $E_G$ is relatively uniform, ~ 1.0 eV, across the surface in the as-grown state. Thermal annealing has essentially no effect on $E_G$ on the e-beam evaporated samples. For 20 Å and 30 Å layers, the average $E_G$ progressively increases, up to ~ 4.0 eV in the latter case, but again annealing has no significant effect. We attribute this different behavior under thermal annealing to the presence of an additional type of atomic defect in the sputtered as opposed to the e-beam evaporated layers. X-ray photoemission spectroscopy (XPS) data on similarly grown layers [8] show a significant, ~ 0.5 eV, shift to lower binding energy (BE) of the Mg and O signals from the oxide for the sputtered MgO in comparison to e-beam deposited layers. Such a shift is consistent with a significant density of negatively charged Mg vacancies in the sputtered oxide. We present STS data below that further support this conclusion, but we note here that the presence of both Mg and O vacancies in the sputtered layers would make them more responsive to moderate thermal annealing than when only O vacancies predominate, as with evaporated MgO. The fact that annealing shifts the XPS peaks to higher BE for the sputtered layers and has minimal effect on the oxide peak locations for the evaporated layers is consistent with this interpretation.

We attribute the increase in $E_G$ that is observed on both types of MgO layers with increasing thickness as arising from a progressive decrease in disorder and defect density in the top layers of the film as it grows in thickness. An Auger and low energy electron diffraction study [9] has shown that MgO films sputter deposited on Ag initially form a distorted rock salt structure, with a 3.6% out of plane expansion for the first several monolayers (ML). In that case the distortion is relaxed after ~10 ML (~22 Å thickness). MgO has a closer lattice match to Ag (2.9%) than Fe (3.7%), so it is reasonable to expect that the thickness required to fully relax the crystal structure in the latter situation would be closer to ~30 Å, the point where we obtain bandgaps close to the bulk crystalline value. A high vacancy-defect density in the thinner MgO layers due to the strong compressive stress at the interface, with the resultant site disorder causing broad band tails and very substantial band-gap narrowing, is certainly consistent with the DOS results reported above; in thicker layers, the vacancy density and oxide disorder decrease as the stress is progressively relaxed away from the interface.

### IV Oxygen and Magnesium Vacancies

Insight into the oxide defects that reduce $E_G$ can be found from the detailed STS study of individual defects in oxide layers of intermediate thickness. We have found that ~ 20 Å oxide layers are optimal for this purpose. In the thinner layers the disorder is too strong to permit resolution of individual defects, while in the thicker layers prolonged DOS measurements are difficult due to poor electronic coupling to the underlying electrode, commonly resulting in an inability of the oxide to withstand the significant tunnel currents required for the measurement without local electronic degradation. However, occasionally in the 20 Å layers, particularly after annealing, it is possible to perform DOS measurements in the vicinity of individual defects. We accomplish this by taking a DOS map at a high bias voltage, and identifying regions, typically ~ 3 nm in extent, of noticeably lower signal amplitude. By placing the STM tip in this region and taking DOS measurements as the tip slowly drifts about in this region, due to the limitations of this room temperature measurement, signals indicative of somewhat isolated defects could sometimes be obtained. An example is given in Fig. 3a, where we show successive DOS sweeps taken while the STM tips drifts about within the circled region of the low spot of the DOS map shown in the inset. There are two distinct, albeit broad, peaks in the DOS, a small peak centered at $V_t$ ~ -0.5V and a larger peak centered at $V_t$ ~ -1.75V, about 1 eV below the apparent, local conduction-band minimum (CBM) of the oxide layer, with the amplitudes of these peaks varying as the tip drifts closer or further away from the defect center. A total-energy functional (TEF) calculation [10] indicates that the neutral oxygen vacancy or F center in MgO should have its filled electron energy levels ~ 1.2 eV below the MgO conduction band. Yet such a vacancy defect, when located within electron tunneling distance of an underlying electrode, can be expected to be in the $F^{2+}$ state at equilibrium if the F state energy is indeed near the CBM. The two peaks then would correspond to the initiation of electron tunneling to the lowest unoccupied level of the defects, resulting in final states $F^+$ and F respectively. The TEF calculation treats the oxide in isolation and the $F^+$ defect with a change in the bound charge density of the six adjacent Mg atoms, producing a level only 0.13 eV below that of the neutral defect. However, metallic screening from the underlying electrode can be expected play a significant role in our experiment which may explain the > 1 eV difference in the positions of the two peaks that we observe. In any case, while the exact identification of the origin of the two DOS peaks that we observe at various locations on the annealed, sputtered MgO layers is at this point unresolved, the origin is almost certainly oxygen vacancies.

We have also observed defect states when they are not so dilute as to be discrete. In Fig. 3b we show a series of DOS sweeps taken with the STM tip over a region of an unannealed 10 Å Mg/10 Å MgO layer where the high-bias DOS map indicated a more bulk-like behavior than average. There we see a broad plateau of band-tail states extending from ~ 0.5 to 2.0 + eV, depending on tip location. It is straightforward to attribute this behavior to the presence of a higher density of defects than in the annealed sample of Fig. 3a; a density sufficient to broaden the peak structure into a distinct defect band below the main conduction band onset, but not sufficient to have the defect states dominant the overall behavior – the case over much of the area of such samples (see Fig. 2a). Note also the substantial band tailing in the valence band in Fig. 3b, due to the defect induced disorder, in contrast to the much lower VBM in the less defective region of the annealed sample shown in Fig. 3a. The role of these defect states in producing broad band-tails and narrowing $E_G$ in the thinnest MgO layers now appears clear.

Previously we had mentioned that XPS measures considerable negative charging in sputtered MgO layers, which we have attributed to the presence of a substantial density of Mg vacancies, V centers. If the electron levels of V centers are close to the VBM [10] they should be occupied due to tunneling from the underlying electrode and will be negatively charged, $V^{2-}$. We have found STS evidence for such defect states in a 30 Å, sputtered and annealed MgO layer. DOS sweeps taken, again while the tip drifts across the surface, show (Fig. 3c) clear evidence of a local peak in the DOS at about 1 eV above the VBM, with data taken at different places over the surface showing peaks at different tip bias points but always in essentially the same position relative to the local VBM.

We expect a considerable impact of these vacancy defects, F centers in evaporated layers and F and V centers in sputtered layers, and the resultant band tails and narrow $E_G$, on the electron transport behavior of MgO MTJs, particularly in the ultra-thin limit. The greatly reduced bandgap, particularly in the thinner layers can explain the low barrier heights obtained from fits to tunneling *I-V* characteristics. Elastic scattering from the atomic defects during tunneling can open up parallel co-tunneling channels that short-circuit the coherent tunneling that otherwise leads, in the ideal limit, to extremely high TMR in Fe(alloy) – MgO – Fe(alloy) (001) junctions [1,2]. Enhanced co-tunneling due to the increased defect density that is observed in the thinner MgO layers is consistent with the degradation of TMR generally found in MTJs with very thin MgO barriers.

We do need to note here that XPS shows that the surface of as-grown and annealed MgO layers is invariably covered by a chemisorbed oxygen layer, presumably bound to the oxide by the positively charged O vacancies. However, this chemisorbed oxygen is not seen by the STM unless measurements are made either at a sufficiently high tunnel current or for a sufficiently long time that the oxide is degraded and the chemisorbed oxygen is stabilized by the locally increased density of oxide defects on the surface into nm sized clusters that can be imaged by the STM [11,12]. Both the behavior of this chemisorbed layer upon the deposition of the top electrode and during annealing of the completed MTJ can have a major impact on the electronic structure of the MgO layer as revealed by the STS measurements reported here. But to the extent that vacancy centers and the resultant disorder remain in the oxide of the completed junction they will play a major role in determining the TMR behavior of the device.

## V Conclusion

We have employed STS to measure the DOS, the local bandgap, and defect energy levels of textured (001) MgO layers grown on (001) Fe electrodes by e-beam evaporation and magnetron sputtering. In the latter case, the deposition process appears to result in a significant density of both O and Mg vacancies in the oxide layers, which makes the MgO amenable to improvement by thermal annealing. This leads to a sufficiently low defect density that, in localized areas, spectroscopic information regarding defect energy levels can be obtained. Two broad defect levels in the upper half of the oxide band gap have been observed that are tentatively assigned to the oxygen vacancy while a defect level ~ 1 eV above the VBM is assigned to the Mg vacancy. In e-beam evaporated and unannealed sputtered films the defect density is sufficiently high that individual defect states cannot be resolved and the resulting disorder yields broad, and rather symmetric band tails that in the thinner 15 Å films extend to within 0.5 eV of $E_f$, consistent with barrier height determinations from I-V measurements on fully formed MTJs. The presence of these vacancy defects is attributed to the compressive strain from the lattice mismatch between the MgO and the Fe electrode, an attribution that is consistent with the defect density decreasing and the bandgap widening with oxide thickness, with the latter approaching the bulk value at ~ 30 Å. Elastic scattering from such vacancy defects can lead to a reduction of TMR in MgO junctions by providing co-tunneling channels in parallel to the coherent tunneling that yields the ideal TMR behavior.

**Acknowledgements**

This research was supported by NSF through the Cornell Center for Materials Research, by the Office of Naval Research and by the ARO/MURI program. The research also benefited from support from NSF through use of the facilities of the Center for Nanoscale Systems and the Cornell Nanoscale Facility/NNIN.

**Figure Captions**

Fig. 1: (Color online) STM topograph (a) and simultaneously acquired DOS map (b) for a 15 Å sputtered, annealed MgO film on Fe, acquired at a tip bias of 2V, 100 pA. There is little overall correlation between electronic structure and topography. DOS measurements (dI/dV) across surface (c - dotted lines) show uniform band gap values (~1.5 V); structure in DOS map is due to variation in strength of band tails. Curves are offset for clarity. Solid curve is from an identical, but unannealed sample. The band gap is (~0.5 V) smaller.

Fig, 2: (Color online) DOS measurements for several 20 Å film preparation methods (a). Sputtering produces a band gap of ~1.5 eV, which annealing further increases to ~2.5 eV. Annealing sputtered films broadens the band gap and reduces vacancy-defect sites. Egun evaporation results in band gaps that are not improved upon annealing. Measurements for 30 Å (b) films show that as the surface moves away from the strained, disordered Fe/MgO interface the vacancy-defect density drops and the band gap increases.

Fig. 3: (Color Online) DOS measurements (a) for a 20 Å annealed, sputter deposited MgO film, taken at the lowest points in the DOS map (inset, circled) show peaks characteristic of oxygen vacancy sites 1.75 V and 0.5 V above the Fermi level. Image parameters were -2.5 V tip bias and 100 pA. (b) shows a peak from a similar DOS spot in an unannealed 20 Å film. The discrete levels become smeared out into a band tail in the more disordered unannealed sample.
(c) DOS measurements on a defective site from a 30 Å sputtered, annealed MgO film. Curves are offset for clarity. Initial curve exhibits very close to the full 7.8 eV MgO bulk band gap, while a defect state (see arrows) due to a Mg vacancy appears in subsequent measurements as the STM tip drifts.   This state is ~1V above the valence band maxima (VBM)

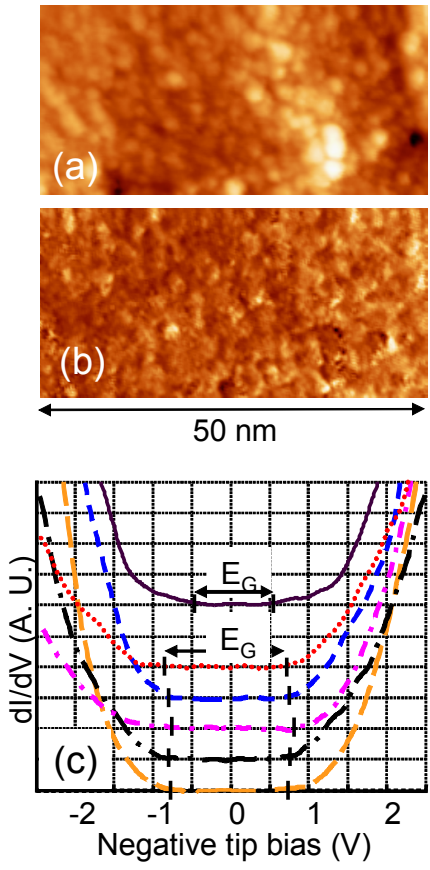

**Figure 1, Mather *et al.***

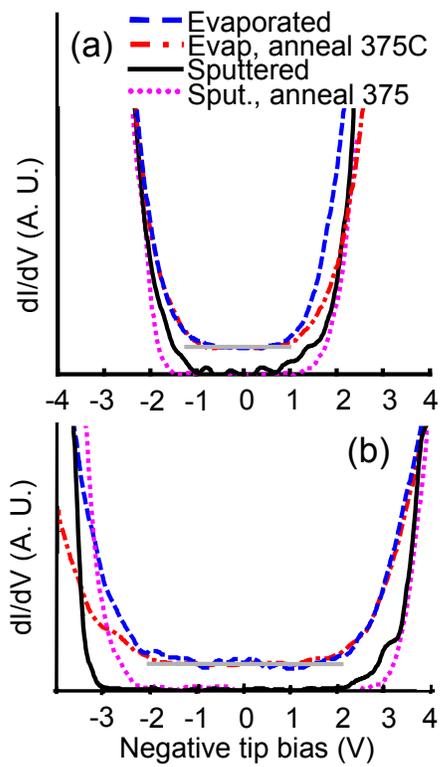

**Figure 2, Mather *et al.***

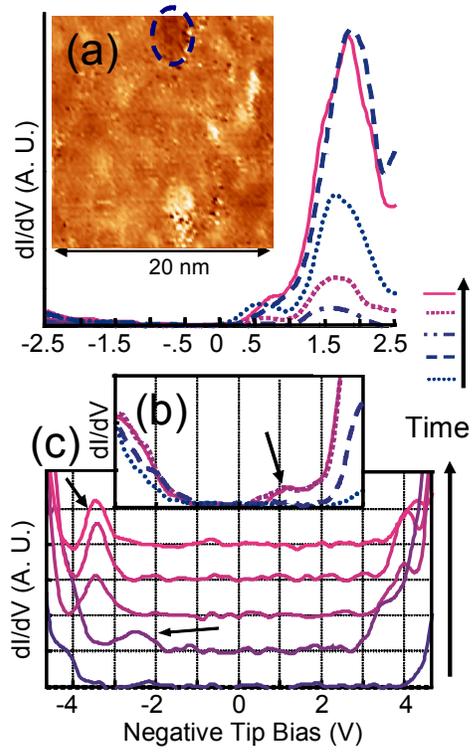

**Figure 3, Mather *et al.***